\UseRawInputEncoding
\documentclass[superscriptaddress, notitlepage, reprint]{revtex4-1}
\usepackage[english]{babel}
\usepackage{amsmath,amsthm}
\usepackage{amsfonts}
\usepackage[pdfborder={0 0 0}, colorlinks=true, urlcolor=blue, linkcolor=blue, citecolor=blue]{hyperref}
\usepackage{color}
\usepackage{graphicx}
\usepackage{float}
\usepackage{subfigure}
\usepackage{lipsum}
\usepackage{epstopdf}
\usepackage{dcolumn}
\begin{document}

\title{Exceptional point in self-consistent Markovian master equations }
\author{Dong  Xie}
\email{xiedong@mail.ustc.edu.cn}
\affiliation{College of Science, Guilin University of Aerospace Technology, Guilin, Guangxi 541004, People's Republic of China}
\author{Chunling Xu}
\affiliation{College of Science, Guilin University of Aerospace Technology, Guilin, Guangxi 541004, People's Republic of China}

\begin{abstract}
 Exceptional point (EP) denotes the non-Hermitian degeneracy,  in which both eigenvalues and eigenstates become identical. By the conventional local Markovian master equation, EP can be constructed by parity-time (PT) or anti-PT symmetry in a system composed of coupled subsystems. However, the coupling between two systems makes the conventional local Markovian master equation become inconsistent. By using self-consistent Markovian master equation, we show that there is no EP in the system composed of two bosonic subsystems. The conventional local master equation can be valid only when the coupling strength is much smaller than the difference in resonance frequency between the two subsystems. In a system composed of three bosonic subsystems, EP can be obtained by adiabatically eliminating one of the three subsystems.
\end{abstract}
\maketitle

\section{Introduction}
The Hamiltonian governing the evolution of the closed system is Hermitian, and thus only degeneracy of the energy levels is possible. The inevitable coupling to the surrounding environment makes the effective Hamiltonian of the open system become non-Hermitian. The non-Hermitian degeneracy, known as EP~\cite{lab1,lab2}, denotes that both eigenvalues and eigenstates coalesce. EPs have recently attracted more and more research, mainly by finding a large number of meaningful applications and exotic phenomena such as loss-induced lasing~\cite{lab3}, stopped light\cite{lab3a}, quantum state control~\cite{lab4,lab5}, asymmetric backscattering~\cite{lab6}, asymmetric mode switching~\cite{lab7,lab8}, energy transfer~\cite{lab9}, phase accumulation~\cite{lab10,lab11}, enhancement of Quantum Heat Engine~\cite{lab12}. More importantly, EPs have been found to play an important role in improving the measurement sensitivity~\cite{lab13,lab14,lab15,lab16,lab17,lab18,lab19,lab20}.

EPs can appear in PT and anti-PT symmetrical systems. In the presence of PT symmetry~\cite{lab21,lab22}, non-Hermitian Hamiltonians can have entirely real eigenvalues. EPs are the separate points between purely real eigenvalues and the normally complex eigenvalues. Similarly, EPs are the separate points between purely imaginary eigenvalues and the normally complex eigenvalues of anti-PT symmetrical non-Hermitian Hamiltonians~\cite{lab23,lab24}.

PT and anti-PT symmetrical non-Hermitian effective Hamiltonian can be constructed by the coupled bosonic systems suffering from the local Markovian dissipation or driving~\cite{lab25}. By transforming the conventional Lindblad master equation into the quantum Heisenberg-Langevin equation, the effective PT and anti-PT symmetrical non-Hermitian effective Hamiltonian for the evolution of bosonic modes can be obtained. However, the local master equation may fail when there is coupling between the systems. It has been shown that local master equation may violate the second law of thermodynamics~\cite{lab26} and give rise to non-physical results~\cite{lab27,lab28,lab29,lab30}, even in the limit of small bath couplings. Recently, it is shown that local master equation may fail to describe dissipative critical behavior~\cite{lab31}. It is therefore necessary to be careful about the coupling between systems when constructing EPs.

Taking into account light-matter interaction, a self-consistent nonlocal Markovian master equation in the dressed picture has been proposed~\cite{lab32,lab33}.
 Recently, self-consistent nonlocal Markovian dissipation master equation for open quadratic quantum system has been derived~\cite{lab34}. In this article, we further derive the self-consistent nonlocal Markovian driving master equation for constructing the PT symmetrical system. We show that a fermionic bath with a strong enough chemical potential is required to obtaining the incoherent driving. By transforming the self-consistent equation into the corresponding quantum Heisenberg Langevin equation, we prove that EP can not appear in the system composed of two subsystems. The conventional local Markovian master equation is reasonable, requiring that not only the coupling strength is much smaller than the bare resonance frequency difference but also that the baths are symmetric. Finally, we show that adiabatically eliminating one of the three coupled subsystems can construct EPs.

This article is organized as follows. In Section II, we introduce the  EPs in PT symmetrical non-Hermitian Hamiltonian by the conventional local Markovian master equation. In Section III, the self-consistent Markovian master equation for general system is reviewed. In Section IV, we obtain the dressed Markovian master equation for the general quadratic system and the condition of incoherent driving is proposed. In Section V, we show that
there is no EP in the system composed of two bosonic subsystems. In Section IV, EP can be obtained by adiabatically eliminating one of the three coupled subsystems. A simple summary and the possibility of experimentation are proposed in Section VII.

\section{EPs by the conventional local Markovian master equation}
\begin{figure}[h]
\includegraphics[scale=0.35]{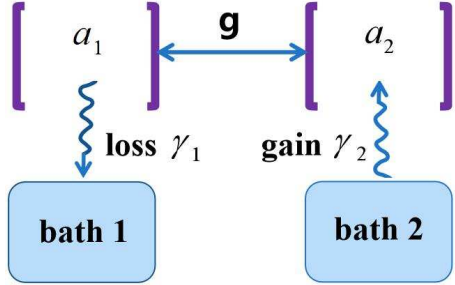}
 \caption{\label{fig.1}Schematic diagram of non-Hermitian Hamiltonian. Two cavity modes are coupled with the strength $g$. Cavity mode 1 suffers from the dissipation bath 1 with the loss rate $\gamma_1$. Cavity mode 2 is incoherent driven by the bath 2 with the gain rate $\gamma_2$. }
\end{figure}

A typical non-Hermitian system is composed of coupled cavities with two resonant modes $a_1$ and $a_2$, as shown in Fig.~\ref{fig.1}, with the non-Hermitian Hamiltonian (setting $\hbar=1$)
\begin{align}
H=(\omega_1-i \gamma_1) a_1^\dagger a_1+(\omega_2+i \gamma_2) a_2^\dagger a_2+g(a_1^\dagger a_2+a_2^\dagger a_1),
\tag{1}
\label{eq:A01}
\end{align}
where $\omega_1$ and $\omega_2$ are the resonance frequencies of the modes 1 and 2, respectively; $\gamma_1$ and $\gamma_2$ are the total loss/gain rates of the modes 1 and 2, respectively.

When resonance frequencies are tuned to be equal (i.e. $\omega_1=\omega_2=\omega$) and the gain rate is equal to the loss rate (i.e. $\gamma_1=\gamma_2=\gamma$), the non-Hermitian Hamiltonian is PT symmetrical. The eigenvalues of the non-Hermitian Hamiltonian $H$ are given by
\begin{align}
E_\pm=\omega\pm\sqrt{g^2-\gamma^2}.
\tag{2}
\label{eq:A02}
\end{align}
When $g=\gamma$, the eigenvalues and the eigenstates are degenerate. Hence, $g=\gamma$ represents the EP.

The non-Hermitian Hamiltonian comes from the conventional local Markovian master equation~\cite{lab25}
\begin{align}
\frac{d\rho}{dt}=-i[H_S,\rho]+\gamma\mathcal{L}(a_1)\rho+\gamma \mathcal{L}(a_2^\dagger)\rho,
\tag{3}
\label{eq:A03}
\end{align}
where the superoperator $\mathcal{L}(a)\rho=2a\rho a^\dagger-a^\dagger a\rho-\rho a^\dagger a$ with $a=\{a_1,\ a_2\}$, and the Hamiltonian $H_S=\omega_1 a_1^\dagger a_1+\omega_2 a_2^\dagger a_2+g(a_1^\dagger a_2+a_2^\dagger a_1).$
The quantum Langenvin equation can be achieved by the formula~\cite{lab35,lab36,lab37}
\begin{align}
\frac{da}{dt}=i[H_S,a]-[a,a_1^\dagger](\gamma a_1-\sqrt{2 \gamma}a_{1\textmd{in}})+(\gamma a_1^\dagger-\sqrt{2 \gamma}a^\dagger_{1\textmd{in}})\nonumber\\
[a,a_1]-[a,a_2](\gamma a_2^\dagger-\sqrt{2 \gamma}a^\dagger_{2\textmd{in}})+(\gamma a_2-\sqrt{2 \gamma}a_{2\textmd{in}})[a,a_2^\dagger],
\tag{4}
\label{eq:4}
\end{align}
where the noise operators satisfy that
\begin{align}
&\langle a_{j\textmd{in}}\rangle=0,\ \ \ \ \ \langle a_{j\textmd{in}}a_{k\textmd{in}}\rangle=0,\tag{5}\\
&\langle a^\dagger_{j\textmd{in}}a_{k\textmd{in}}\rangle=0,\langle a_{j\textmd{in}}a^\dagger_{k\textmd{in}}\rangle=\delta_{jk}.
\tag{6}
\label{eq:A06}
\end{align}

\section{self-consistent Markovian master equation}
However, the master equation in Eq.~(\ref{eq:A03}), is not self-consistent due to the coupling between the two subsystems. A self-consistent master equation can be derived by the dressed master equation~\cite{lab32,lab33,lab34,lab38}
\begin{align}
\frac{d\rho}{dt}=-i[H_S+H_{LS},\rho]+\mathcal{D}[\rho],\tag{7}
\label{eq:A07}
\end{align}
where $H_{LS}=\sum_{\alpha,\beta,\omega}L_{\alpha\beta}(\omega)O^\dagger_\alpha(\omega)O_\beta(\omega)$ is a Lamb-shift correction, and the superoperator $\mathcal{D}[\rho]$ is described by
\begin{align}
\mathcal{D}[\rho]=\sum_{\alpha,\beta,\omega}\lambda_{\alpha\beta}(\omega)[2O_\beta(\omega)\rho O_\alpha^\dagger(\omega)-O_\beta(\omega) O_\alpha^\dagger(\omega)\rho-\nonumber\\
\rho O_\beta(\omega) O_\alpha^\dagger(\omega)].\tag{8}
\label{eq:A08}
\end{align}
Here, the dressed operator is given by
\begin{align}
O_\alpha(\omega)=\sum_{k,q}\delta_{\omega_q-\omega_k}|k\rangle\langle k|Q_\alpha|q\rangle\langle q|,\tag{9}
\label{eq:A09}
\end{align}
and the factors are
\begin{align}
\lambda_{\alpha\beta}(\omega)&=\frac{1}{2}\int_{-\infty}^\infty d\tau e^{i \omega \tau}\langle {\tilde{R}}^\dagger_\alpha(\tau)R_\beta\rangle,\tag{10}\label{eq:10}\\
L_{\alpha\beta}(\omega)&=\frac{1}{2i}\int_{0}^\infty d\tau [e^{i \omega \tau}\langle {\tilde{R}}^\dagger_\alpha(\tau)R_\beta\rangle-e^{-i \omega \tau}\langle {\tilde{R}}^\dagger_\alpha(\tau)R_\beta\rangle],\tag{11}
\label{eq:A11}
\end{align}
where $\omega_k$ and $|k\rangle$ are the $k$th eigenvalue and eigenstate of the system Hamiltonian $H_S$, the undressed operator $O_\alpha$ acts on the system in the interaction Hamiltonian $H_{\textmd{int}}=\sum_\alpha O_\alpha\bigotimes R_\alpha$ ($R_\alpha$ acts on the environment), $\langle.\rangle$ denotes the expectation value calculated with the environment density operator $\rho_E$, and $\tilde{R}_\alpha(\tau)=e^{iH_E\tau}{R}_\alpha e^{-iH_E\tau}$ with the environment Hamiltonian $H_E$.
\section{dressed master equation for general quadratic system}
For a general quadratic bosonic system, the Hamiltonian is described by~\cite{lab39}
\begin{align}
H'=\sum_{n=1}^NH_n+\sum_{i=1,i<j}^NH_{ij},\tag{12}
\label{eq:12}
\end{align}
in which,

\begin{align}
&H_n=\omega_na^\dagger_n a_n+(\frac{\chi_n}{2}a_n^2+h.c.),\tag{13}\\
&H_{ij}=(g_{ij}a_ia_j+\lambda_{ij}a_ia_j^\dagger)+h.c.,\tag{14}
\label{eq:14}
\end{align}
where $H_{n}$ denotes the Hamiltonian for the $n$th subsystem  with $n=\{1,...,N\}$, $\omega_n$ is the resonance frequency of the bosonic
subsystem with the annihilation operator $a$ and the creation operator $a^\dagger$, $|\chi_n|$ denotes the strength of two-photon driving, and $\lambda_{ij}$ ($g_{ij}$) denotes the coupling strength of the rotating (counterrotating)-wave  interaction between the two subsystems.

By using an Hopfield-Bogoliubov (HB) transformation~\cite{lab40,lab41}, for the stable normal phase, the total Hamiltonian can be rewritten as a diagonal form
\begin{align}
H'=\sum_{n=1}^N\Omega_n b_n^\dagger b_n+\frac{1}{2}(\Omega_n-\omega_n),\tag{15}
\label{eq:15}
\end{align}
where the collective bosonic mode operators $b_n$ satisfy the commutation relation: $[b_i,b_j^\dagger]=\delta_{ij}$,
\begin{align}
b_n=\sum_{i=1}^N(\mu_{ni}a_{i}+\nu_{ni}a_{i}^\dagger)/\xi_n,\tag{16}
\end{align}
in which, the normalization factor is described by $\xi_n=\sqrt{\sum_{i=1}^N(|\mu_{ni}|^2-|\nu_{ni}|^2)}$.
In the Nambu space, it can be rewritten as
\begin{align}
\vec{\mathbf{b}}=\mathbb{T} \vec{\mathbf{a}},\tag{17}
\end{align}
where the Nambu field vector is defined as
\begin{align}
\vec{\mathbf{x}}=(x_1,...,x_N,x^\dagger_1,...,x^\dagger_N),\tag{18}
\end{align}
 and the canonical transformation matrix
 \[
 \mathbb{T}= \left(
\begin{array}{ll}
\mathbb{\tilde{\mu}}\ \ \ \mathbf{\tilde{\nu}}\\
\mathbf{\tilde{\nu}}^* \ \ \mathbf{\tilde{\mu}}^*\\
  \end{array}
\right ) \tag{19} ,\label{eq:19}\]
where the elements of the matrix are $\tilde{\mu}_{ni}=\mu_{ni}/\xi_n$ and $\tilde{\nu}_{ni}=\nu_{ni}/\xi_n$.
The coefficient vectors $(\mu_{n1},...,\mu_{nN},\nu_{n1},...,\nu_{nN})^T$ are eigenvectors of the HB matrix \textbf{M}, which is derived by the commutation relation $[b_n,H']=\Omega_nb_n$~\cite{lab39}
 \[
 \mathbf{M}= \left(
\begin{array}{ll}
\mathbf{A}\ \ \ -\mathbf{B}\\
\mathbf{B}^* \ \ -\mathbf{A}^*\\
  \end{array}
\right )  ,\tag{20}\label{eq:20}\]
with submatrixs
 \[
 \mathbf{A}= \left(
\begin{array}{ll}
\omega_1\ \ \lambda_{12}\ \ldots\lambda_{1N}\\
\lambda_{12}^*\ \omega_2\ \ldots\  \lambda_{2N}\\
\ \vdots\ \ \ \ \ \vdots\ \ \ \ddots\ \ \ \vdots\\
\lambda_{1N}^*\ \lambda_{2N}^*\ \ldots\omega_N
 \tag{21}\label{eq:21} \end{array}
\right )  ,\]
 \[
 \mathbf{B}= \left(
\begin{array}{ll}
\chi_1\ \ g_{12}\ \ldots\ g_{1N}\\
g_{12}\ \ \chi_2\ \ldots\  g_{2N}\\
 \ \vdots\ \ \ \ \ \ \ \vdots\ \ \ddots\ \ \ \vdots\\
 g_{1N}\ g_{2N}\ \ldots\ \chi_N
  \end{array}
\right )  .\tag{22}\label{eq:22}\]

\subsection{Incoherent dissipation and driving}
For the dissipation environment, there are $N$ independent thermal baths with the Hamiltonian described by

\begin{align}
H_{E,n}=\int dk (\epsilon_n(k)-\eta_n)c_n^\dagger(k)c_n(k),\tag{23}
\end{align}
where the spectrum $\epsilon_n(k)\geq0$ is non-negative and $\eta_n$ denotes the chemical potential of the $n$th thermal bath at temperature $T_n$.

The bath operators $c_n(k)$ satisfy the rules
\begin{align}
&\{c_n(k),c^\dagger_n(q)\}_{\zeta_n}=\delta(k-q),\tag{24}\\
&\{c_n(k),c_n(q)\}_{\zeta_n}=0,\ [c_n(k),c_m(q)]=0,\tag{25}
\end{align}
where $\{X,Y\}_{\zeta_n}=XY+\zeta_n YX$, and $\zeta_n=+1\ (-1)$ belongs to fermionic (bosonic) systems.
The density matrix of thermal bath can be described by
\begin{align}
\rho_E=\bigotimes_{n=1}^N\frac{e^{-H_{E,n}/T_n}}{\textmd{Tr}(e^{-H_{E,n}/T_n})}\tag{26},
\end{align}
Based on the above equations, the two-point expectation values can be obtained
\begin{align}
&\langle c_n(k)c_m(q)\rangle=0,\tag{27}\\
&\langle c^\dagger_n(k)c_m(q)\rangle=\delta_{mn}\delta(k-q)f_n(\epsilon_n(k)),\tag{28}
\end{align}
in which,
\begin{align}
f_n(\epsilon)=[\zeta_n+e^{(\epsilon-\eta_n)/T_n}]^{-1}\tag{29}.
\end{align}

We consider that $N$ subsystems interact linearly with the corresponding bath separately, which are described by
\begin{align}
H_{\textmd{int}}&=\sum_{n=1}^N(a_n+a_n^\dagger)\otimes\int dk g_n(k)[c_n(k)+c^\dagger_n(k)]\nonumber\\
&=\sum_{n=1}^N O_n\otimes R_n.\tag{30}
\end{align}
In the basis of the Hamiltonian $H_S$, the eigenoperator associated with $O_n$ can be obtained
\begin{align}
O_n(\omega)=\sum_{k=1}^N[\phi_{n,k}\delta_{\omega,\Omega_k}b_k+\phi_{n,k}^*\delta_{\omega,-\Omega_k}b^\dagger_k],\tag{31}
\end{align}
where the element
$\phi_{n,k}=(\mathbb{T}^{-1})_{n,k}+(\mathbb{T}^{-1})^*_{n,k+N}$.

The factor $\lambda_{nm}$ in Eq.~(\ref{eq:10}) can be derived by the correlation functions~\cite{lab34}
\begin{align}
&\lambda_{nm}(\omega)=\delta_{nm}\lambda_{nn}(\omega), \tag{32}
\end{align}
 \[
 \lambda_{nn}(\omega)=\left\{
\begin{array}{ll}
\mathcal{J}_n(\omega)[1-\zeta_n f_n(\omega)]\ \ \ \ \ \ \ \ \ \ &\textmd{if}\ \  \omega>0\\
\mathcal{J}_n(-\omega)f_n(-\omega)\ \ \ \ \ \ \ \ \ \ \ \ \ \ \ &\textmd{if}\ \ \omega<0\\
\mathcal{J}_n(0)[1+(1-\zeta_n) f_n(0)]\ \ \ &\textmd{if}\ \  \omega=0
  \end{array}\right\}
  ,\tag{33}\label{eq:33}\]

where the spectral density for the $n$th bath is given by
\begin{align}
\mathcal{J}_n(\omega)=\pi\int dk|g_n(k)|^2\delta(\omega-\epsilon_n(k)).\tag{34}\label{eq:34}
\end{align}

In a general case, the system is not degenerate (i.e. $\Omega_n\neq\Omega_m$ when $n\neq m$) and the eigenspectrum is not zero (i.e. $\Omega_n\neq0$),
the superoperator $\mathcal{D}(\rho)$  can be described by
\begin{align}
\mathcal{D}(\rho)=\sum_{n,k}\gamma_{n,k}[[1-\zeta_n f_n(\Omega_k)]\mathcal{L}(b_k)\rho+f_n(\Omega_k)\mathcal{L}(b^\dagger_k)\rho],\tag{35}\label{eq:35}
\end{align}
where the coupling constants $\gamma_{n,k}=\mathcal{J}_n(\Omega_k)|\phi_{n,k}|^2$.

Combining Eq.~(\ref{eq:4}) and Eq.~(\ref{eq:35}), we can see that the total change rate of the $n$th mode $b_k$ is
\begin{align}
\Gamma_k=\sum_n\Gamma_{k,n}=\sum_n\gamma_{n,k}[1-\zeta_n f_n(\Omega_k)-f_n(\Omega_k)].\tag{36}\label{eq:36}
\end{align}
When all the baths are composed of bosons, the total change rate $\Gamma_k=\sum_n\gamma_{n,k}>0$ denotes that the mode is suffering from the dissipation process.

In order to obtain the incoherent gain, several baths must be composed of fermions, i.e., $\zeta_n=1$. In this case, the change rate $\Gamma_{k,n}=\gamma_{n,k}[1-2f_n(\Omega_k)]$ should be negative. As a result, it leads to that $\Omega_k<\eta_n$, which means that a strong enough chemical potential is needed to drive the system incoherently. It can also be considered that the bath is composed of spins, which are in the excited states. In other words, bosonic bath is not suitable for implementing incoherent driving.

\section{EP does not exist in two-boson systems}
In this section, we investigate whether EP can exist in the system composed of two bosonic subsystems with the self-consistent Markovian master equation.

The bosonic system composed of two subsystems is dominated by the Hamiltonian
\begin{align}
H_{S1}=\omega a_1^\dagger a_1+\omega a_2^\dagger a_2+g(a_1^\dagger a_2+a_2^\dagger a_1)\tag{37}\label{eq:37}.
\end{align}
The subsystem 1 interacts with the bath 1 composed of bosons at zero temperature, and the subsystem 2 interacts with the bath 2 composed of fermions at zero temperature. The corresponding interaction Hamiltonian is described by
\begin{align}
H_{\textmd{int1}}=\sum_{n=1}^2(a_n+a_n^\dagger)\otimes\int dk g_n(k)[c_n(k)+c^\dagger_n(k)].\tag{38}\label{eq:38}
\end{align}

The Hamiltonian can be diagonalized as
\begin{align}
H_{{S}1}=(\omega+g)b_1^\dagger b_1 +(\omega-g)b_2^\dagger b_2,\tag{39}\label{eq:39}
\end{align}
where the dressed operators are $b_1=(a_1+a_2)/\sqrt{2}$ and $b_2=(a_1-a_2)/\sqrt{2}$.
Then, using the consistent master equation in Eq.~(\ref{eq:35}) for $g\neq0$, we can achieve
\begin{align}
\frac{d\rho}{dt}=&-i[H_{S1},\rho]+[{J}_1(\omega+g)\mathcal{L}(b_1)\rho+{J}_2(\omega+g)\mathcal{L}(b_1^\dagger)\rho\nonumber\\
&+{J}_1(\omega-g)\mathcal{L}(b_2)\rho+ {J}_2(\omega-g)\mathcal{L}(b_2^\dagger)\rho].\tag{40}\label{eq:40}
\end{align}
We assume that baths have a very large bandwidth, leading to that $\gamma={J}_j(\omega+g)={J}_j(\omega-g)$ with $j=1,2$ and the Lambshift correction $H_{{LS}1}=0$.
The quantum-Heisenberg Langevin equation are given by using Eq.~({\ref{eq:4}})
\[
i \left(
\begin{array}{ll}
\dot{b}_1\\
\dot{b}_2\\
  \end{array}
\right )= \left(
\begin{array}{ll}
\omega+g\ \ \ \ 0\\
\ \ \ 0 \ \ \ \ \omega-g\\
  \end{array}
\right ) \left(
\begin{array}{ll}
{b}_1\\
{b}_2\\
  \end{array}
\right )+ \sqrt{2\gamma}\left(
\begin{array}{ll}
b_{1\textmd{in}}+b^\dagger_{1\textmd{in}}\\
b_{2\textmd{in}}+b^\dagger_{2\textmd{in}}\\
  \end{array}
\right )
,\tag{41}\label{eq:41}\]
where $b_{\textmd{in}}$ denotes the noise operator.
The effective Hamiltonian is described by
\[
H_{\textmd{eff}1}= \left(
\begin{array}{ll}
\omega+g\ \ \ \ 0\\
\ \ \ 0 \ \ \ \ \omega-g\\
  \end{array}
\right ).\tag{42}\label{eq:42}\]
Due to that $H_{\textmd{eff}1}$ is Hermitian, EP does not exist. This result shows that EP can not be constructed in a resonance-coupled driven-dissipative system. Because of the resonant coupling, both the driving and the dissipation act synchronously on each subsystem. The asymmetric effects of driving and dissipation can not be obtained.
\subsection{The self-consistent master equations with degenerate eigenvalues}
In the non-degenerate case, the dressed modes $b_k$ are independent, leading to that EP does not exist.  Then, we consider the degenerate eigenenergies of the system Hamiltonian.

The system possesses $M$ different energy eigenspaces, labeled by an index $\iota=1,...,M$. There are $N_\iota$ eigenvectors  for eigenvalue $\omega_\iota$.
The consistent Markovian master equation for the baths with large bandwidth can be given by~\cite{lab34}
\begin{align}
\frac{d\rho}{dt}=-i[H_S,\rho]+\sum_{n,\iota}\sum_{\alpha=1,\beta=1}^{N_\iota}[\Phi_{\mu\nu}^{(n,\iota)}\lambda_{nn}(\omega_\iota)(2b_\mu\rho b^\dagger_\nu - \nonumber\\ \{b^\dagger_\nu b_\mu,\rho\}_+)+\Phi_{\nu\mu}^{(n,\iota)}\lambda_{nn}(-\omega_\iota)(2b^\dagger_\mu\rho b_\nu-\{b_\nu b^\dagger_\mu,\rho\}_+)],
\tag{43}\label{eq:43}
\end{align}
where the factors $\Phi_{\mu\nu}^{(n,\iota)}=\phi_{n,\mu}\phi^*_{n,\nu}$.

A simple degenerate case is one in which there is only paring coupling between two bosonic subsystems with Hamiltonian
\begin{align}
H_{S2}=\omega a_1^\dagger a_1+\omega a_2^\dagger a_2+g(a_1 a_2+a_1^\dagger a_2^\dagger),
\tag{44}\label{eq:44}
\end{align}
The subsystem 1 interacts with the bosonic bath 1 at zero temperature, and the subsystem 2 interacts with the fermionic bath 2 at zero temperature.

The transformation matrix between Nambu field vector $\vec{\mathbf{a}}=(a_1,a_2,a_1^\dagger,a_2^\dagger)$ and $\vec{\mathbf{b}}=(b_1,b_2,b_1^\dagger,b_2^\dagger)$ can be expressed as

 \[
 \mathbb{T}^{-1}= \left(
\begin{array}{ll}
W_+\ \ 0 \ \ \ \ \ 0 \ \ \ W_-\\
\ 0\ \  \ W_+\ \ W_-\ \ 0\\
\ 0\ \  \ W_-\ \ W_+\ \ 0\\
W_-\ \ 0 \ \ \ \ \ 0 \ \ \  W_+
  \end{array}
\right),\tag{45}\label{eq:45}\]
where the values are defined as $W_\pm=\pm\sqrt{{\frac{\omega}{2\sqrt{\omega^2-g^2}}\pm\frac{1}{2}}}$.

In the rotating reference frame, we can obtain the dynamic of the expectation values of the dressed operators $b_1,\ b_2$ according to Eq.~(\ref{eq:43})
\[
i \left(
\begin{array}{ll}
\dot{\langle b_1\rangle}\\
\dot{\langle b_2\rangle}\\
  \end{array}
\right )= \left(
\begin{array}{ll}
\frac{-i(W\lambda_-+\lambda_+)}{2}\ \ \ \frac{-i\sqrt{W^2-1}\lambda_-}{2}\\
\frac{-i\sqrt{W^2-1}\lambda_-}{2}\ \ \ \ \frac{-i(W \lambda_--\lambda_+)}{2}\\
  \end{array}
\right ) \left(
\begin{array}{ll}
\langle b_1\rangle\\
\langle b_2\rangle\\
  \end{array}
\right ),\tag{46}\label{eq:46}\]
where the factors are defined as $W=\frac{\omega}{2\sqrt{\omega^2-g^2}}$ and $\lambda_\pm=\mathcal{J}_1(\sqrt{\omega^2-g^2})\pm\mathcal{J}_2(\sqrt{\omega^2-g^2})$.
Therefore, the effective Hamiltonian $H_{\textmd{eff}2}$ can be expressed as
\[
H_{\textmd{eff}2}= \left(
\begin{array}{ll}
\frac{-i(W\lambda_-+\lambda_+)}{2}\ \ \ \frac{-i\sqrt{W^2-1}\lambda_-}{2}\\
\frac{-i\sqrt{W^2-1}\lambda_-}{2}\ \ \ \ \frac{-i(W \lambda_--\lambda_+)}{2}\\
  \end{array}
\right ).\tag{47}\label{eq:47}\]
The eigenvalues of $H_{\textmd{eff}2}$ can be derived, which are given by
\begin{align}
E_\pm=\frac{i}{2}[\sqrt{\lambda_+^2+(W^2-1)\lambda_-^2}\pm W\lambda_-].
\tag{48}\label{eq:48}
\end{align}
Due to that the eigenvalues are still imaginary, there are no EPs that separate the real and imaginary values.
Therefore, in the degenerate eigenspace, EPs are not allowed in the self-consistent Markovian master equation.

As a summary, EPs can not appear in the driven-dissipative bosonic system irrespective of whether the eigenvalues of the eigensystem are degenerate or not.
\section{EPs by adiabatic elimination}
In this section, we try to construct the EP by adiabatic elimination in multiple boson systems.

Firstly, we find the condition that the conventional local Markovian master equation can be close to the nonlocal self-consistent master equation.
For two non-resonant coupled subsystems, the Hamiltonian is described by
\begin{align}
H_{S3}=\omega_1 a_1^\dagger a_1+\omega_2 a_2^\dagger a_2+g(a_1^\dagger a_2+a_2^\dagger a_1),
\tag{49}\label{eq:49}
\end{align}
In the diagonalized form, the Hamiltonian is rewritten as $H_{S3}=\Omega_+b_1^\dagger b_1+\Omega_-b_2^\dagger b_2$
with the eigenvalues $\Omega_\pm=\frac{\omega_1+\omega_2\pm\sqrt{4g^2+\Delta^2}}{2}$.
The canonical transformation matrix is given by
 \[
 \mathbb{T}^{-1}= \left(
\begin{array}{ll}
\sqrt{\frac{1+Y}{2}}\  \ -\sqrt{\frac{1-Y}{2}}\ \ \ \ \ \ 0 \ \ \ \ \ \ \ \ \ \ \ 0\\
\sqrt{\frac{1-Y}{2}}\ \ \ \ \sqrt{\frac{1+Y}{2}}\ \ \ \ \ \ \ \ 0 \ \ \ \ \ \ \ \ \ \ \ 0\\
\ \ \ 0\ \ \ \ \ \ \ \ \ \ \ \ 0  \ \ \ \ \ \ \ \ \sqrt{\frac{1+Y}{2}}\  \ \  -\sqrt{\frac{1-Y}{2}}\\
\ \ \ 0\ \ \ \ \ \ \ \ \ \ \ \ 0\  \ \ \ \ \ \ \ \sqrt{\frac{1-Y}{2}}\  \ \ \ \sqrt{\frac{1+Y}{2}}
  \end{array}
\right ),\tag{50}\label{eq:50}\]
where the value $Y$ is defined as $Y=\frac{\Delta}{\sqrt{\Delta^2+4g^2}}$, and the resonance frequency difference $\Delta$ is given by $\Delta=\omega_1-\omega_2$. Without loss of generality, next we consider that both the resonance frequency difference and the coupling strength are larger than 0, i.e., $\Delta>0$ and $g>0$.

When the coupling strength is much less than the resonance frequency difference ($g\ll\Delta$), we can obtain that
\begin{align}
a_1&\simeq b_1-\frac{g}{\Delta} b_2+O(\frac{g^2}{\Delta^2})b_1+O(\frac{g^3}{\Delta^3})b_2, \tag{51}\\
a_2&\simeq b_2+\frac{g}{\Delta} b_1+O(\frac{g^2}{\Delta^2})b_2+O(\frac{g^3}{\Delta^3})b_1,\tag{52}\\
\Omega_1&\simeq\omega_1+\frac{g^2}{\Delta}+O(\frac{g^3}{\Delta^3}),\tag{53}\\
\Omega_2&\simeq\omega_2-\frac{g^2}{\Delta}-O(\frac{g^3}{\Delta^3}),\tag{54}\\
 \gamma_{21}&=\gamma_{12}\simeq O(\frac{g^2}{\Delta^2}),\tag{55}
\end{align}
 where $O(\frac{g^2}{\Delta^2})$ denotes the second order small quantity, and $O(\frac{g^3}{\Delta^3})$ denotes the third order small quantity. Ignoring all the small quantities, the local Markovian master equation is recovered
\begin{align}
&\frac{d\rho}{dt}\approx\sum_{k=1}^2\{-i[\omega_k a_k^\dagger a_k,\rho]+\mathcal{L}_{\textmd{loc}}(\rho)\},\tag{56}\label{eq:56}\\
&\textmd{in which},\nonumber\\
 &\mathcal{L}_{\textmd{loc}}(\rho)=
\gamma_{k,k}[[1-\zeta_k f_k(\omega_k)]\mathcal{L}(a_k)\rho+f_k(\omega_k)\mathcal{L}(a^\dagger_k)\rho].\tag{57}
\end{align}
In the above local Markovian master equation, the two bosonic modes $a_k$ are independent. Therefore, EPs can not occur in the local Markovian master equation.

Up to the first order small quantity (i.e. on the order of $g/\Delta$), the self-consistent Markovian master equation can be given by
\begin{align}
&\frac{d\rho}{dt}\approx-i[H_{S3},\rho]+\mathcal{L}_{loc}(\rho)+\frac{g}{\Delta}\sum_{k=1}^2(-1)^{k-1}\gamma_{k,k}\{[1-\nonumber\\
&\zeta_k f_k(\omega_k)](2a_1\rho a_2^\dagger+2a_2\rho a_1^\dagger)+f_k(\omega_k)(2a_1^\dagger\rho a_2+2a_2^\dagger\rho a_1)\nonumber\\
&-[1+(1-\zeta_k )f_k(\omega_k)](a_1^\dagger a_2+a_2^\dagger a_1)\rho+\rho(a_1^\dagger a_2+a_2^\dagger a_1)\}.\tag{58}
\end{align}
 When the two baths are identical and their temperatures are close to zero degrees ($\zeta_1 f_1(\omega_1)=\zeta_2 f_2(\omega_2)$), and the couplings between the subsystems and the corresponding baths are the same ($\gamma_{11}=\gamma_{22}$), we can achieve the conventional Markovian master equation from the above equation
 \begin{align}
\frac{d\rho}{dt}\approx-i[H_{S3},\rho]+\mathcal{L}_{loc}(\rho).\tag{59}
\end{align}
It shows that the conventional local Markovian master equation is reasonable, requiring that not only the coupling strength $g$ is much smaller than the resonance frequency difference $\Delta$ (rather than the resonance frequencies) but also that the baths are symmetric.

Then, we consider a system composed of three bosonic subsystems, with the Hamiltonian described by
\begin{align}
H_{S4}=\sum_{i=1}^3\omega_i a_i^\dagger a_i+g(a_1^\dagger a_3+a_3^\dagger a_1)+g'(a_2^\dagger a_3+a_3^\dagger a_2),\tag{60}
\label{eq:60}
\end{align}
where $\omega_i$ denotes the frequencies of the $i$th subsystem, and $g$ ($g'$) denotes the coupling strength between modes 1 (2) and 3.

In the rotating reference frame, the Hamiltonian $H_{S4}$ can be  rewritten as
\begin{align}
H_{S4}=&\Delta' a_1^\dagger a_1+(\Delta'-\varepsilon) a_2^\dagger a_2+g(a_1^\dagger a_3+a_3^\dagger a_1)\nonumber\\
&+g'(a_2^\dagger a_3+a_3^\dagger a_2),\tag{61}
\label{eq:61}
\end{align}
where $\Delta'=\omega_1-\omega_3$ and $\varepsilon=\omega_1-\omega_2$.

We consider that the condition satisfies $\Delta'\gg g$, so that the self-consistent Markovian master equation is close to the local Markovian master equation according to Eq.~(\ref{eq:56}). Then, we obtain the unitary transformation matrix $U$ defined in $(b_1,b_2,b_3)^{T}=U(a_1,a_2,a_3)^{T}$ for $\varepsilon \simeq O(g)$, which is close to
 \[
 U\approx \left(
\begin{array}{ll}
\ 1\ \  \ \ \ \frac{g}{\Delta'}\ \ \frac{g}{\Delta'}\\
\frac{-g}{\Delta'}\ \ \ \ 1\ \ \ \ \frac{g}{\Delta'}\\
\frac{-g}{\Delta'}\ \ \ \frac{-g}{\Delta'} \ \ \ 1
  \end{array}
\right ).\tag{62}
\label{eq:62}\]

When the temperature in all three baths is zero, the self-consistent Markovian master equation can give the Langevin equation with the diagonalized modes
\[
i \left(
\begin{array}{ll}
\dot{\langle b_1\rangle}\\
\dot{\langle b_2\rangle}\\
\dot{\langle b_3\rangle}
  \end{array}
\right )\approx- \left(
\begin{array}{ll}
\Upsilon_1\ 0 \ \ 0\\
0\ \ \Upsilon_2 \ 0\\
0\ \ \ 0 \ \ \Gamma_2
  \end{array}
\right ) \left(
\begin{array}{ll}
\langle b_1\rangle\\
\langle b_2\rangle\\
\langle b_3\rangle
  \end{array}
\right ),\tag{63}\]
where the values are defined as $\Upsilon_1=\Gamma_1+i\Delta'$, $\Upsilon_2=\Gamma_2+i(\Delta'-\varepsilon)$, $\Gamma_1=\mathcal{J}_1(\omega_1)=\Gamma_2=\mathcal{J}_2(\omega_2)$, and $\Gamma_3=\mathcal{J}_3(\omega_3)$.
Based on the above eqaution, the evolution of modes $a_j$ can be derived by
\[
i \left(
\begin{array}{ll}
\dot{\langle a_1\rangle}\\
\dot{\langle a_2\rangle}\\
\dot{\langle a_3\rangle}
  \end{array}
\right )\approx-U^{-1}\left(
\begin{array}{ll}
\Upsilon_1\ 0 \ \ 0\\
0\ \ \Upsilon_2 \ 0\\
0\ \ \ 0 \ \ \Gamma_2
  \end{array}
\right ) U \left(
\begin{array}{ll}
\langle a_1\rangle\\
\langle a_2\rangle\\
\langle a_3\rangle
  \end{array}
\right ),\tag{64}\]\\
After calculation from the above equation, we can achieve the detail evolution dynamic
\[
i \left(
\begin{array}{ll}
\dot{\langle a_1\rangle}\\
\dot{\langle a_2\rangle}\\
\dot{\langle a_3\rangle}
  \end{array}
\right )\approx-\Lambda \left(
\begin{array}{ll}
\langle a_1\rangle\\
\langle a_2\rangle\\
\langle a_3\rangle
  \end{array}
\right ),\tag{65}\label{eq:65}\]\\
where the evolution matrix is given by
\[\Lambda=
\left(
\begin{array}{ll}
\ \ \ \ \Upsilon'\  \ \ \ \ \frac{g^2 (\Gamma_3 -\Gamma_1)}{\Delta'^2}\ \frac{g (\Gamma_1-\Gamma_3)}{\Delta'}\\
\frac{g^2 (\Gamma_3 - \Gamma_1)}{\Delta'^2}\ \ \ \ \Upsilon'\ \ \ \ \ \ \frac{g (\Gamma_1- \Gamma_3)}{\Delta'}\\
 \frac{g (\Gamma_1 -\Gamma_2)}{\Delta'}\ \frac{g (\Gamma_1 - \Gamma_3)}{\Delta'}\ \ \ \ \ \ \Gamma_3
  \end{array}
\right ),\tag{66}\]
in which, $\Upsilon'=\Gamma_1+\frac{g^2 (2 \Gamma_1 + \Gamma_3)}{\Delta'^2}$.
 When $\Gamma_2\gg \Gamma_1\gg\Delta'$, the mode $\langle a_3\rangle$ can be eliminated due to that the mode $\langle a_3\rangle$ reaches the steady state much faster than the other two modes.  By assuming $\langle \dot{a}_3\rangle=0$, we can obtain
 \begin{align}
\langle a_3\rangle\approx\frac{g}{\Delta'\Gamma_2}(\Gamma_2-\Gamma_1)(\langle a_1\rangle+\langle a_2\rangle),
\tag{67}
\label{eq:67}
\end{align}
 Using the above equation, we can obtain the evolution equation for the modes 1 and 2
 \[
i \left(
\begin{array}{ll}
\dot{\langle a_1\rangle}\\
\dot{\langle a_2\rangle}
  \end{array}
\right )=H_{\textmd{eff}3}\left(
\begin{array}{ll}
\langle a_1\rangle\\
\langle a_2\rangle
  \end{array}
\right ),\tag{68}\]
where the effective Hamiltonian of the reduced two subsystems is depicted as
 \[H_{\textmd{eff}3}=i\left(
\begin{array}{ll}
-\Gamma_1-i\Delta'\ \ \  \ \ -\frac{g^2\Gamma_1}{\Delta'^2}\\
\ \  -\frac{g^2\Gamma_1}{\Delta'^2}\ \ \  \ \ -\Gamma_1-i(\Delta'-\varepsilon)
  \end{array}
\right ).\tag{69}\]

Moving to the reference frame rotating with frequency $\Delta'-\varepsilon/2$, the effective Hamiltonian can be reduced to
\[H_{\textmd{eff}3}=i\left(
\begin{array}{ll}
-\Gamma_1-i\varepsilon/2\ \ \  -\frac{g^2\Gamma_1}{\Delta'^2}\\
\ \ -\frac{g^2\Gamma_1}{\Delta'^2}\ \ \  \ \ -\Gamma_1+i\varepsilon/2
  \end{array}
\right ).\tag{70}\]
The eigenvalues of $H_{\textmd{eff}3}$ are given by $\Omega_\pm=i(-\Gamma_1\pm\sqrt{g^4\Gamma_1^2/\Delta'^4-\varepsilon^2/4})$. As a consequence, the EP appears at $\varepsilon=2g^2\Gamma_1/\Delta'^2$, which separates the purely imaginary eigenvalue and the normally complex eigenvalue. Therefore, EP appears in the effective anti-PT symmetrical Hamiltonian.

Then, we try to construct the EP without the condition $g\ll \Delta'$, i.e., there is no need to approach the local Markovian master equation.  We redefine that $\varepsilon=2\Delta'$. In this case, the unitary transformation matrix $U$ can be exactly derived, which is described by
 \[
 U= \left(
\begin{array}{ll}
\frac{\Delta'+\Delta_g}{2\Delta_g}\ \ \frac{g^2}{\Delta_g(\Delta_g+\Delta')}\ \ \frac{g}{\Delta_g}\\
\frac{\Delta'-\Delta_g}{2\Delta_g}\ \ \frac{g^2}{\Delta_g(\Delta'-\Delta_g)}\ \ \frac{g}{\Delta_g}\\
\ \frac{-g\Delta_g}{\Delta'^2}\ \ \ \ \ \ \frac{g\Delta_g}{\Delta'^2}\ \ \ \ \ \ \ \frac{\Delta_g}{\Delta'}
  \end{array}
\right ),\tag{71}\]
where $\Delta_g=\sqrt{\Delta'^2+2g^2}$.
In this case, the evolution matrix $\Lambda$ defined in Eq.~({\ref{eq:65}}) can be reformulated as
 \[\Lambda=U^{-1}\left(
\begin{array}{ll}
\Gamma_1+i\Delta'\ \ \ \ \ \ \ 0 \ \ \ \ \ \ \ \ \ \ \ 0\\
\ \ \ \ \ 0\ \ \ \ \ \ \ \ \Gamma_1-i\Delta'\ \ \ \ \ 0\\
\ \ \ \ \ 0\ \ \ \ \ \ \ \ \ \ \ \ 0 \ \ \ \ \ \ \ \ \ \ \ \Gamma_3
  \end{array}
\right )U=\]
\[\left(
\begin{array}{ll}
\ \ \ \ \ \frac{Z_-}{\Delta_g^2}\ \ \ \ \ \ \ \ \ \ \ \ \ \ \ \frac{i g^2(\Gamma_1-\Gamma_3)}{\Delta_g^2}\ \ \ \ \ \frac{i g \Delta'(\Gamma_3-\Gamma_1)-g\Delta_g^2}{\Delta_g^2}\\
\ \frac{i g^2(\Gamma_1-\Gamma_3)}{\Delta_g^2}\ \ \ \ \ \ \ \ \ \ \ \ \ \ \ \frac{Z_+}{\Delta_g^2}\ \  \ \ \ \ \ \ \ \frac{i g \Delta'(\Gamma_1-\Gamma_3)-g\Delta_g^2}{\Delta_g^2}\\
 \frac{i g \Delta'(\Gamma_3-\Gamma_1)-g\Delta_g^2}{\Delta_g^2}\ \frac{i g \Delta'(\Gamma_1-\Gamma_3)-g\Delta_g^2}{\Delta_g^2}\ \frac{-i(2g^2\Gamma_1+\Gamma_3\Delta'^2)}{\Delta_g^2}
  \end{array}
\right ),\tag{72}\]
where $Z_\pm=i[g^2(\Gamma_1+\Gamma_2)+\Gamma_1\Delta'^2]\pm\Delta'\Delta_g^2$.
For $|\Gamma_3|\gg|\Gamma_1| \textmd{and} \ |\Delta'|$, we can adiabatically eliminate the mode $a_3$. As a result, the effective Hamiltonian for modes $a_1$ and $a_2$ can be described by
\[H_{\textmd{eff4}}=\left(
\begin{array}{ll}
{\Gamma_2\Delta'\Delta_g^2-i\kappa}\
\ \ \ \ \ \ -i \chi\\
 \ \ \ \ \ -i \chi\ \ \ \ \ \ \ {-\Gamma_2\Delta'\Delta_g^2-i\kappa}
  \end{array}
\right ),\tag{73}\]
where
$\kappa=\Delta'^2(g^2+\Gamma_1\Gamma_3)+g^2(2g^2+\Gamma_1^2+\Gamma_1\Gamma_3)$ and $\chi=g^2(\Delta_g^2+\Gamma_1^2-\Gamma_1\Gamma_3)$.
The eigenvalues of $H_{\textmd{eff4}}$ are given by $E_\pm=-i\kappa\pm\sqrt{(\Gamma_3\Delta'\Delta_g^2)^2-\chi^2}$.
When the modes 1, 2 and 3 are suffering from incoherent dissipation, i.e., $\Gamma_1>0$ and $\Gamma_3>0$, $|\chi|=\Gamma_3\Delta'\Delta_g^2$ denotes the anti-PT symmetrical EP. Therefore, the relation $\Gamma_1\approx\frac{\Delta'(\Delta'^2+2g^2)}{g^2}$ is required to find the anti-PT symmetrical EP.

In order to find the PT symmetrical EP, the condition $\kappa=0$ is necessary. The modes 1 and 2 are incoherent driven, i.e., $\Gamma_1<0$.
As a result, we can achieve that $\Gamma_2=\frac{g^2(2g^2+\Delta'^2+\Gamma_1^2)}{(g^2+\Delta'^2)|\Gamma_1|}$. The PT symmetrical EP can also appear at $\Gamma_1\approx\frac{\Delta'(\Delta'^2+2g^2)}{g^2}$.

\section{conclusion }
We investigate the construction of EPs in the effective non-Hermitian Hamiltonian from the self-consistent Markovian master equation. Unlike the result from the conventional local Markovian master equation, we prove that the EPs can not exist in the system composed of two bosons. For constructing the PT symmetrical system, we further derive the self-consistent nonlocal Markovian driving master equation. We show that fermionic bath with a strong enough chemical potential is required to obtaining incoherent driving. It can also be considered that the bath is composed of spins, which are in the excited states by extra controls. Bosonic bath is not suitable for implementing incoherent driving. And we show that the conventional local Markovian master equation is reasonable, which requires not only that the coupling strength is much smaller than the resonance frequency difference (rather than the resonance frequencies) but also that the baths are symmetrical. By adiabatically eliminating one of the three coupled subsystems, we can reconstruct the EPs with two different parameter choices: one is that the coupling strength is much less than the resonance frequency difference, and the other is that the coupling strength is not much less than the resonance frequency difference. The former can only construct anti-PT symmetrical EPs. The latter can construct anti-PT and PT symmetrical EPs.

The system composed of three coupled bosonic subsystems in this article can be realized in magnon-cavity-magnon coupled system~\cite{lab24} or in a variety of different photonic and phononic systems\cite{lab42}. Our work lays the foundation for constructing real EPs in non-Hermitian systems.

\section*{Acknowledgements}
This research was supported by the National Natural Science Foundation of China under Grant No. 62001134, Guangxi Natural Science Foundation under Grant No. 2020GXNSFAA159047.


\begin{thebibliography}{9}

\vspace{3mm}

\bibitem{lab1}M.-A. Miri and A. Al\'{u}, Exceptional points in optics and photonics, Science 363, 42 (2019).
\bibitem{lab2}\c{S}. K. \"{O}zdemir, S. Rotter, F. Nori, and L. Yang, Parity-time symmetry and exceptional points in photonics, Nat. Mater. 18, 783 (2019).
\bibitem{lab3}B. Peng, \c{S}. K. \"{O}zdemir, S. Rotter, H. Yilmaz, M. Liertzer, F. Monifi, C. M. Bender, F. Nori, and L. Yang, Loss-induced suppression and revival of lasing, Science 346, 328 (2014).
\bibitem{lab3a}Tamar Goldzak, Alexei A. Mailybaev, and Nimrod Moiseyev, Light Stops at Exceptional Points, Phys. Rev. Lett. 120, 013901 (2018).
\bibitem{lab4}W. Liu, Y. Wu, C.-K. Duan, X. Rong, and J. Du, Dynamically Encircling an Exceptional Point in a Real Quantum System, Phys. Rev. Lett. 126, 170506 (2021).
\bibitem{lab5}M. Abbasi, W. Chen, M. Naghiloo, Y. N. Joglekar, and K. W. Murch, Topological quantum state control through exceptional-point proximity, Phys. Rev. Lett. 128. 160401 (2022).
\bibitem{lab6}S. Soleymani, Q. Zhong, M. Mokim, S. Rotter, R. ElGanainy, and \c{S}. K. \"{O}zdemir, Chiral and degenerate perfect absorption on exceptional surfaces, Nat. Commun. 13, 599 (2022).
\bibitem{lab7}A. Li, J. Dong, J. Wang, Z. Cheng, J. S. Ho, D. Zhang, J. Wen, X.-L. Zhang, C. T. Chan, A. Al¨², C.-W. Qiu, and L. Chen, Hamiltonian Hopping for Efficient Chiral Mode Switching in Encircling Exceptional Points, Phys. Rev. Lett. 125, 187403 (2020).
\bibitem{lab8}J. W. Yoon, Y. Choi, C. Hahn, G. Kim, S. Ho Song, K.-Y. Yang, J. Yub Lee, Y. Kim, C. S. Lee, J. K. Shin, H.-S. Lee, and P. Berini, Time-asymmetric loop around an exceptional point over the full optical communications band, Nature (London) 562, 86 (2018).
\bibitem{lab9}H. Xu, D. Mason, L. Jiang, and J. G. E. Harris, Topological energy transfer in an optomechanical system with exceptional points, Nature (London) 537, 80 (2016).
 \bibitem{lab10}T. Gao, E. Estrecho, K. Y. Bliokh, T. C. H. Liew, M. D. Fraser, S. Brodbeck, M. Kamp, C. Schneider, S. H\"{o}fling, Y. Yamamoto, F. Nori, Y. S. Kivshar, A. G. Truscott, R. G. Dall, and E. A. Ostrovskaya, Observation of non-Hermitian degeneracies in a chaotic exciton-polariton billiard, Nature (London)/, 526, 554 (2015).
\bibitem{lab11}M. S. Ergoktas, S. Soleymani, N. Kakenov, K. Wang, T. B. Smith, G. Bakan, S. Balci, A. Principi, K. S. Novoselov,
\c{S}. K. \"{O}zdemir, and C. Kocabas, Topological engineering of terahertz light using electrically tunable exceptional point singularities, Science 376, 184 (2022).
\bibitem{lab12}J.-T. Bu, J.-Q. Zhang, G.-Y. Ding, J.-C. Li, J.-W. Zhang, B. Wang, W.-Q. Ding, W.-F. Yuan, L. Chen, \c{S}. K. \"{O}zdemir, F. Zhou, H. Jing, and M. Feng, Enhancement of Quantum Heat Engine by Encircling a Liouvillian Exceptional Point, Phys. Rev. Lett. 130, 110402 (2023).
\bibitem{lab13}J. Wiersig, Enhancing the sensitivity of frequency and energy splitting detection by using exceptional points: Application to microcavity sensors for single-particle detection. Phys. Rev. Lett. 112, 203901 (2014).
\bibitem{lab14}J. Wiersig,  Sensors operating at exceptional points: General theory, Phys. Rev. A 93, 033809 (2016).
\bibitem{lab15}W. Chen,  \c{S}. K. \"{O}zdemir, G. Zhao, J. Wiersig, L. Yang,  Exceptional points enhance
sensing in an optical micro-cavity. Nature 548, 192 (2017).
\bibitem{lab16}Hossein Hodaei, Absar U. Hassan, Steffen Wittek, Hipolito Garcia-Gracia, Ramy El-Ganainy, Demetrios N. Christodoulides, Mercedeh Khajavikhan,
Enhanced sensitivity at higher-order exceptional points, Nature 548, 187 (2017).
\bibitem{lab17}P-Y. Chen, M. Sakhdari, M. Hajizadegan, Q. Cui, MM-C. Cheng, R. El-Ganainy, A. Al\`{u}
Generalized parity-time symmetry condition for enhanced sensor telemetry, Nat. Electron 1, 297 (2018).
\bibitem{lab18}Z. Dong, Z. Li, F. Yang, C-W. Qiu, J. S. Ho, Sensitive readout of implantable
microsensors using a wireless system locked to an exceptional point. Nat Electron
2, 335 (2019).
\bibitem{lab19}D. Xie, C. Xu, A. Wang, Parameter estimation and quantum entanglement in PT symmetrical cavity magnonics system, Results Phys. 26, 104430 (2021).
\bibitem{lab20}J. M. P. Nair, D. Mukhopadhyay, and G. S. Agarwal, Enhanced Sensing of Weak Anharmonicities through Coherences in Dissipatively
Coupled Anti-PT Symmetric Systems, Phys. Rev. Lett. 126, 180401 (2021).
\bibitem{lab21}C. M. Bender and S. Boettcher, Real Spectra in NonHermitian Hamiltonians Having PT Symmetry, Phys.
Rev. Lett. 80, 5243 (1998).
\bibitem{lab22}C. M. Bender, Making sense of non-Hermitian Hamiltonians, Rep. Prog. Phys. 70, 947 (2007).
\bibitem{lab23}F. Yang, Y.-C. Liu, and L. You, Anti-PT symmetry in dissipatively coupled optical systems,
Phys. Rev. A 96, 053845 (2017).
\bibitem{lab24}J. Zhao, Y. Liu, L. Wu, C. Duan, Y. Liu, and J. Du, Observation of Anti-PT -Symmetry Phase Transition in the
Magnon-Cavity-Magnon Coupled System, Phys. Rev. Applied. 13, 014053 (2020).
\bibitem{lab25}M. Zhang, W. Sweeney, C. W. Hsu, L. Yang, A. D. Stone, and L. Jiang,  Quantum Noise Theory of Exceptional Point Amplifying Sensors, Phys. Rev. Lett. 123, 180501 (2019).
\bibitem{lab26}A. Levy and R. Kosloff, The local approach to quantum transport may violate the second law of thermodynamics, Europhys.
Lett. 107, 20004 (2014)
\bibitem{lab27}M. T. Naseem, A. Xuereb, O. E. Mustecaplioglu, Thermodynamic consistency of the optomechanical master equation,
Phys. Rev. A 98, 052123 (2018).
\bibitem{lab28}G. De Chiara, G. Landi, A. Hewgill, B. Reid, A. Ferraro, A. J. Roncaglia and M. Antezza, Reconciliation of quantum
local master equations with thermodynamics, New J. Phys. 20,
113024 (2018).
\bibitem{lab29}M. T. Mitchison and M. B. Plenio, Non-additive dissipation in open quantum networks out of equilibrium, New J. Phys. 20,
033005 (2018).
\bibitem{lab30}M. Cattaneo, G. L. Giorgi, S. Maniscalco, and R. Zambrini, Local versus global master equation with common and separate baths: Superiority of the global approach in partial secular approximation, New J. Phys. 21, 113045 (2019).
\bibitem{lab31}M. Konopik and E. Lutz, Local master equations may fail to describe dissipative critical behavior, Phys. Rev. Research 4, 013171 (2022).
\bibitem{lab32}F. Beaudoin, J. M. Gambetta, and A. Blais, Dissipation and ultrastrong coupling in circuit QED, Phys. Rev. A 84, 043832 (2011).
\bibitem{lab33}A. Settineri, V. Macr¨ª, A. Ridolfo, O. Di Stefano, A. F. Kockum, F. Nori, and S. Savasta, Dissipation and thermal noise in hybrid quantum systems in the ultrastrong-coupling regime, Phys. Rev. A 98, 053834 (2018).
\bibitem{lab34}A. D'Abbruzzo and D. Rossini, Self-consistent microscopic derivation of Markovian master equations for open quadratic quantum systems, Phys. Rev. A 103, 052209 (2021).
\bibitem{lab35}C. Gardiner, P. Zoller, Qauntum Noise: A Handbook of Markovian and Non-Markovian Quantum Stochastic Methods with Applications to Quantum
Optics, vol. 56 (Springer, Berlin, 2004)
\bibitem{lab36}M. Reitz, C. Sommer, C. Genes, Langevin approach to quantum optics with molecules. Phys. Rev. Lett. 122, 203602 (2019)
\bibitem{lab37}D. Xie, C. Xu, A. Wang, Quantum thermometry with a dissipative quantum Rabi system, Eur. Phys. J. Plus 137, 1323 (2022).
\bibitem{lab38}H.-P. Breuer and F. Petruccione, The Theory of Open Quantum Systems (Oxford University Press, Oxford, 2007).
\bibitem{lab39}D. Xie, C. Xu, A. Wang, Quantum phase transition revealed by the exceptional point in a Hopfield-Bogoliubov matrix, Phys. Rev. A 104, 062418 (2021).
\bibitem{lab40}J. J. Hopfield, Theory of the contribution of excitons to the complex dielectric constant of crystals, Phys. Rev. 112, 1555 (1958).
\bibitem{lab41}C. Ciuti, G. Bastard, and I. Carusotto, Quantum vacuum properties of the intersubband cavity polariton field, Phys. Rev. B 72, 115303 (2005).
Cao Yunshan, Yan Peng. Exceptional magnetic sensitivity of PT-symmetric cavity
magnon polaritons. Phys Rev B 2019;99:214415.
\bibitem{lab42}A. McDonald, T. Pereg-Barnea, and A. A. Clerk, Phase-Dependent Chiral Transport and Effective Non-Hermitian Dynamics
in a Bosonic Kitaev-Majorana Chain, Phys. Rev. X 8, 041031 (2018).
\end{thebibliography}
\end{document}